\documentstyle [fleqn,12pt]{article}
\newcommand{\dy}[2]{\frac{{\textstyle \partial \/} #1}{{\textstyle 
\partial \/} #2}}

\begin{document}
\baselineskip .75cm 
\begin{titlepage}
\title{ \bf Quasi-particle model of QGP - a revisit }      
\author{Vishnu M. Bannur \\
{\it Department of Physics}, \\  
{\it University of Calicut, Kerala-673 635, India.} }
%\date{}
\maketitle
\begin{abstract} 

The quasi-particle model of quark gluon plasma (QGP) is revisited here 
with a new method, different from earlier 
studies, without the need of temperature dependent bag constant  
as well as other effects such as confinement effects, 
effective degrees of freedom etc. Our model has only two 
system dependent parameters and surpraisingly good fit to lattice 
results for gluon plasma, 2-flavor and 3-flavor QGP are obtained. 
The basic idea is to evaluate energy density $\varepsilon$ first from grand 
partition function of quasi-particle QGP and then derive all other 
thermodynamic functions from $\varepsilon$. Quasi-particles are assumed 
to have temperature dependent mass equal to plasma frequency. 
Energy density, pressure and speed of sound at zero chemical potential 
are evaluated and compared with available lattice data. 
We further extend the model to finite 
chemical potential, without any new parameters, to obtain quark density, 
quark susceptibility etc. and fits very well with the lattice results on 
2-flavor QGP.        
 
\end{abstract}
\vspace{1cm}

\noindent
{\bf PACS Nos :} 12.38.Mh, 12.38.Gc, 05.70.Ce, 52.25.Kn \\
{\bf Keywords :} Equation of state, quark gluon plasma, 
quasi-particle plasma  
\end{titlepage}
%%%%%%%%%%%%%
\section{Introduction :}
 
The non-ideal behaviour of QGP seen in lattice simulations 
\cite{ka.1,ka.2,ka.3} of QCD, and the 
elliptical flow observed in relativistic heavy ion collisions (RHICs) 
lead to hot debate on the nature of QGP near the critical temperature, 
$T_c$, for the last few years \cite{qgp.1}. 
Various models such as (a) QGP with 
confinement effects, (b) strongly interacting QGP (sQGP), 
(c) strongly coupled quark gluon plasma (SCQGP), and 
(d) quasi-particle QGP (qQGP). In model (a), confinement effects like 
bag constant \cite{rh.1}, Cornel potential \cite{ba.1} etc. 
are assumed to give rise to 
non-ideal behaviour of QGP. In (b) \cite{sh.1}, effects of bound states of 
colorless as well as colored hadron resonances are assumed to be 
responsible. In (c) \cite{ba.2, ba.3}, it is assumed that QGP 
near $T_c$ is what is called 
strongly coupled plasma (SCP) and EoS of SCP in QED with proper 
modifications for QCD fits very well lattice data. In (d), QGP is made 
up of quasi-particles with temperature dependent mass \cite{pe.1, pe.2}.  
Different varities of qQGP were proposed in order to make the theory 
thermodynamically consistent as well as consistent with perturbative 
and non-perturbative calculations of QCD \cite{lh.1,s.1}. 

Here we revisit qQGP with a new method, which involves a less 
number of parameters, to derive EoS of QGP at zero 
chemical potential ($\mu$). 
Here $\mu$ means quark chemical potential which is 
one third of the baryon chemical potential.  
We start from energy density, rather than pressure as done earlier, 
and derive various thermodynamic properties by fixing the parameters 
of the model. Further we extend the model to include finite chemical 
potential and obtain quark density ($n_q$), 
change in pressure ($\Delta P$) due to finite $\mu$ and quark 
susceptibility for 2-flavor QGP.                 
  
\section{Phenomenological Model :}
 
The basic assumption in this model is that quarks and gluons in QGP 
are not bare quarks and gluons, but they are quasi-particles. Because 
of their collective behaviour in QGP, massless partons acquire masses  
equal to their respective plasma frequencies and become quasi-partons. 
Following the standard procedure of statistical mechanics \cite{pa.1}, 
the grand partition function is given by,  
\begin{equation}
Q_G = \sum_{s,r} e^{- \beta (E_r - \mu N_s)}
\,\,, \end{equation}
where the sum is over energy states $E_r$ and particle number states $N_s$.    
Now we assume that QGP is made up of non-interacting quasi-partons and on 
taking thermodynamic limit, we get,  
\begin{equation} 
q \equiv \ln Q_G = \mp \sum_{k=0}^\infty 
\ln (1 \mp z\, e^{ - \beta \epsilon_k})  
\,\,, \end{equation} 
where $q$ is called q-potential and $\mp$ for bosons and fermions. 
$\beta$ and $z$ are temperature and fugacity respectively. $\epsilon_k$ 
is the single particle energy, given by, 
\[ \epsilon_k = \sqrt{k^2 + m^2 (T,\mu)} \,\,, \]   
where $k$ is momentum and $m^2$ is the temperature dependent mass. 
The main effects of the interaction of bare partons, namely collective 
effects, are taken in ($T$, $\mu$) dependent mass term  
and treat them as 
non-interacting quasi-partons. With this assumption,  
the average energy $U$ is given by, 
\begin{equation}
U \equiv <E_r> = - \frac{\partial }{\partial \beta} \ln Q_G 
= \sum_k \frac{z \,\epsilon_k e^{- \beta \epsilon_k }}{1 \mp 
z\, e^{- \beta \epsilon_k} } 
\,\,,\end{equation}
which on taking continum limit and after some algebra, we get, 
\begin{equation}
\varepsilon = \frac{g_f\, T^4}{2\, \pi ^2} \sum_{l=1}^\infty 
(\pm 1)^{l-1} z^l \frac{1}{l^4} \left[ (\frac{m\,l}{T})^3 K_1 (\frac{m\,l}{T}) 
+  3\, (\frac{m\,l}{T})^2 K_2 (\frac{m\,l}{T}) \right] 
\,\,, \label{eq:u} \end{equation}
where $g_f$ is the degenarcy and equal to $g_g \equiv 16$ for gluons and 
equal to $2\,n_f$ for quarks. $n_f$ is the number of flavors. 
$K_1$ and  $K_2$ are modified Bessel functions 
of order 1 and 2 respectively. 

\section{Thermodynamics ($\varepsilon$, $P$, $C_s^2$) of QGP with zero $\mu$:} 

Let us first consider the EoS of QGP 
with zero chemical potential and take $z=1$. 
Hence we get the energy density, expressed in terms of 
$e(T) \equiv \varepsilon / \varepsilon_s$,  
for the quark gluon plasma of quasi-partons is 
\[ e(T) = \frac{15}{\pi ^4} \frac{1}{(g_f + \frac{21}{2} n_f)} 
\sum_{l=1}^\infty  \frac{1}{l^4}  \left( g_f \,
\left[ (\frac{m_g\,l}{T})^3 K_1 (\frac{m_g\,l}{T}) 
+  3\, (\frac{m_g\,l}{T})^2 K_2 (\frac{m_g\,l}{T}) \right] \right. \]
\begin{equation}
 + 12\,n_f\,  
(-1)^{l-1}  \left. \left[ (\frac{m_q\,l}{T})^3 K_1 (\frac{m_q\,l}{T}) 
+  3\, (\frac{m_q\,l}{T})^2 K_2 (\frac{m_q\,l}{T}) \right] \right)  
\,\,, \end{equation}
where $\varepsilon_s$ is the Stefan-Boltzman gas limit of QGP, which 
may be obtained by taking high temperature limits of Eq. (\ref{eq:u}) 
for gluons and quarks separately and adding them. 
$m_g$ is the temperature dependent gluon mass, which is equal to the 
plasma frequency, i.e,   
$m_g^2 = \omega_p^2 = \frac{g^2 T^2}{18} (2 N_c + n_f)$. 
This is obtained from the finite temperature perturbative calculations
\cite{th.1,bz.1}. Note that both gluons and quarks contributes to 
the gluon mass. 
However, in the case of $m_q$, the temperature dependent mass of quarks, 
we take $m_q^2 = \frac{g^2 T^2}{18} n_f$ and no contribution from 
gluons at linear response level. $g^2$ is related to the 
two-loop order running coupling constant, given by, 
\begin{equation} \alpha_s (T) = \frac{6 \pi}{(33-2 n_f) \ln (T/\Lambda_T)}
\left( 1 - \frac{3 (153 - 19 n_f)}{(33 - 2 n_f)^2}
\frac{\ln (2 \ln (T/\Lambda_T))}{\ln (T/\Lambda_T)}
\right)  \label{eq:ls} \;, \end{equation}
where $\Lambda_T$ is a parameter related to QCD scale parameter. This 
choice of $\alpha_s (T)$ is motivated from lattice simulations.  
With these values of masses with above $\alpha_s$, we 
can evaluate the $e(T)$ from Eq. (\ref{eq:u}). Note that the only 
temperature dependence in $e(T)$ comes from $\alpha_s(T)$, which 
has the same form as that of lattice simulations \cite{ka.1} with 
$\Lambda_T$ as a free parameter.  
Pressure can be calculated from the thermodynamic relation 
$\varepsilon =  T \frac{\partial P}{\partial T} - P $ and we get 
\begin{equation} \frac{P}{T} = \frac{P_0}{T_0} + \int_{T_0}^{T} dT \,
\frac{\varepsilon (T)}{T^2} \,\, , \label{eq:p} \end{equation}
where $P_0$ and $T_0$ are pressure and temperature at 
some reference points.  
Note that the standard relation of pressure and q-potential 
is not valid here because of temperature dependent $\epsilon_k$. 
We can rederive it using the same procedure from Patria \cite{pa.1}, 
as follows, 
\begin{equation}
\delta q = \frac{1}{Q_G} \left[ \sum_{r,s} e^{- \beta (E_r - \mu N_s)}
\, \left( - E_r \,\delta \beta - \beta \,\delta E_r + N_s \,\delta (\beta \mu) 
\,\right) \right] 
\end{equation}
\[ \frac{P\,V}{T} = q + \int d\beta\, \beta \, 
\frac{\partial m}{\partial \beta}\,<\frac{\partial E_r}{\partial m}> \] 
\[ \frac{P\,V}{T} = \mp \sum_{k=0}^\infty \ln (1 \mp z\, 
e^{ - \beta \epsilon_k})  + F(T)\, V \] 
Therefore, the standard relation of pressure and q-potential is 
not thermodynamically consistent in our case. 
One need extra term arising because 
of the temperature dependent mass. Hence we use Eq. (\ref{eq:p}) to 
evaluate pressure. 
Note that earlier works are all based on modeling the extra term 
so as to fit the lattice data. 

Once we know $P$ and $\varepsilon$, 
$c_s^2 = \dy{\textstyle P}{\textstyle \varepsilon}$ can be evaluated.

\section{Thermodynamics ($n_q$, $\Delta P$, $\chi _q$) of QGP 
with finite $\mu$:} 

Recently, there are lot attempts to simulate QCD with finite $\mu$ 
on lattice and results are reported in Ref. \cite{fk.1,ak.1} etc. 
Here we consider the recent results of Allton {\it et. al.} \cite{ak.1} 
for 2-flavor QGP and try to extend our 
model to finite $\mu$ and explain their results. Again using the 
standard procedure of statistical mechanics, we have, 
\begin{equation}
<N> = z \frac{\partial }{\partial z} \ln Q_G  \,\,, 
\end{equation}
which on continum limit and after some algebra, reduces to 
\begin{equation}
 \frac{n_q}{T^3} = \frac{12}{\pi ^2} \sum_{l=1}^\infty 
(-1)^{l-1} \frac{1}{l^3} \left[ (\frac{m_q\,l}{T})^2 K_2 (\frac{m_q\,l}{T}) 
\sinh(\frac{m_q \,l}{T} \right] 
\,\,. \label{eq:nq} \end{equation}
Now we modify earlier $m_q^2 (T)$ to $m_q^2 (T,\mu)$ as 
\begin{equation}
m_q^2 (T,\mu) = \frac{g^2 T^2}{18} n_f ( 1 + \frac{\mu}{\pi^2\, T^2})\,\,, \end{equation} 
inspired by QCD perturbative calculations \cite{ps.1}. 
In our case $n_f = 2$ and $g^2$ 
is related to two-loop order running coupling constant, discussed earlier, 
but need to be modified to take account of finite $\mu$. Following the work of 
Schneider \cite{sch.1} and Letessier, Rafelski \cite{lr.1}, 
now we change $T/\Lambda_T$ in Eq. (\ref{eq:ls}) as 
\begin{equation} 
\frac{T}{\Lambda_T} \,\sqrt{1 + a \,\frac{\mu^2}{T^2} } \,\,,  
\label{eq:lsa} \end{equation} 
where $a$ is a parameter which is equal to $(1.91/2.91)^2$ in the 
calculation of Schneider for $\mu /T \le 1$  
and $1/\pi^2$ in a phenomological model of Letessier and Rafelski.  

From $n_q$, we may obtain other thermodynamic quantities like, 
\begin{equation}
\Delta P \equiv P(T,\mu) - P(T,0) 
= \int_0^\mu n_q d\mu \,\, 
\end{equation} 
and 
\begin{equation}
\chi_q =  \frac{\partial n_q}{\partial \mu}|_{\mu=0} \,\,.
\end{equation}

\section{Results :} 

It is interesting to see that that this simple model very nicely 
fits lattice data \cite{ka.1,ka.2,ka.3} on all three systems, namely, 
gluon plasma, 2-flavor and 3-flavor QGP in the case of 
zero chemical potential.  
In Fig. 1, we ploted $P(T) / T^4$ Vs $T$ for 
pure gauge, 2-flavor and 3-flavor QGP along with lattice results. 
Note that, in the case of flavored QGP, since there is 
$(10\% \pm 5\%)$ uncertainty in $P$ data \cite{ka.3} on taking continum limit 
with massless quarks \cite{ka.3}, we multiply the lattice 
data by the factor $1.1$ and is plotted. For each system 
$\Lambda_T$ are adjusted so that we get a good fit to lattice results. 
We have fixed $P_0$ from the lattice data at the critical temperature $T_c$ 
for each system. Surprisingly good fit is obtained for 
all systems with $\Lambda_T$ equal to 173, 119 
and 74 MeV for gluon plasma, 2-flavor and 3-flavor QGP respectively. 
We have taken $n_f$ equal to $0$, $2$ and $3$ respectively for three systems. 

Once $P(T)$ is obtained, then other macroscopic quantities 
may be derivable from $P(T)$ and no other 
parameters are needed. In Fig. 2, we plotted $\varepsilon / T^4$ 
Vs $T/ T_c$ for all three systems along with lattice 
results \cite{ka.4} and it fits well without any extra parameters. 
All the three curves looks similar, but shifts to left 
as flavor content increases. We have taken $T_c$ equal to $275$, $175$ and 
$155$ $MeV$ respectively for gluon plasma, 2-flavor and 3-flavor QGP.  

In Fig. 3,  $c_s^2$ is plotted for all three systems, again with 
lattice results for gluon plasma. Reasonably good fit for gluon plasma 
and our model's out come for the flavored QGP. All the three curves have 
similar behaviour, i.e, sharp rise near $T_c$ and then flatten to the 
value close to $1/3$. $c_s^2$ is larger for larger flavor content. For 
2-flavor QGP it is almost coinsides with that of gluon plasma.   

In Fig. 4, $n_q /T^3$ is plotted for 2-flavor QGP and compared with recent 
lattice data with out any new parameters. Only the parameter needed is 
$\Lambda_T$ which is fixed by the results of QGP with zero $\mu$ earlier. 
For different values of $\mu /T_c$, all curves have similar behaviour 
and very nicely fits with lattice points for $T \ge 1.2 \,T_c$. Note that 
this result is for $a = (1.91/2.91)^2$ \cite{sch.1} in Eq. (\ref{eq:lsa}). 
For the other value, $a= 1/ \pi^2$ \cite{lr.1}, results are not satisfactory, 
eventhough both results coinside for $T > 1.5\,T_c$.  

$\Delta P / T^4$ and $\chi_q /T^2$ are plotted in Fig. 5 and 6 
and again fits with the lattice results. Note that in the case finite 
$\mu$ case also we multiply the lattice data by a factor $1.1$ as we 
did for $\mu =0$ case. 

Very close to $T=T_c$, i.e., $T < 1.2 \,T_c$,    
fits or outcome  of our model is not good, especially for 
$\varepsilon$, $c_s^2$, $n_q$ etc. 
Lattice data also has large error bars 
very close to $T_c$. However, except for small region at 
$T=T_c$, our results are very good for all regions of $T > T_c$. 
For $T < T_c$, any way our model is not applicable because 
the system may not be in plasma state. 

\section{Conclusions :}

We revisited the quasi-particle picture of QGP with a 
new method with less parameters.  In this method we derive 
the energy density, $\varepsilon (T)$, first and {\it not} 
$P(T)$ as done earlier, and derive $P(T)$, $c_s^2$ etc.  
QGP is assumed to be consists of non-interacting 
quasi-partons with temperature and chemical potential dependent 
masses. Interactions between 
bare partons lead to collective effects in QGP and hence they 
acquire masses equal to plasma frequencies. The plasma frequency 
depends on running coupling constant and we used 2-loop order 
$\alpha_s(T,\mu)$, which is similar to that of lattice simulations, 
with one parameter related to QCD scale parameter. 
The thermodynamic properties of QGP depends on this parameter 
as well as one integration constant in pressure term which we fix 
to the lattice point at $T_c$.  
Both the parameters are different for different systems like 
gluon plasma, 2-flavor and 
3-flavor QGP etc. Using two system dependent parameter, very 
good fits to lattice results were obtained for 
energy density, pressure, speed of sound, quark density, 
quark susceptibility etc.
No other effects like temperature dependent 
bag pressure, confinement effects, effective degrees of freedom etc. 
were needed to fit the lattice 
results. In comparison with other models, which generally use more 
than two system dependent paarameters, here, just with two system 
dependent parameter and using $\alpha_s(T,\mu)$, 
inspired by lattice simulation of QCD, fits lattice results very nicely. 
Hence the non-ideal effects seen in lattice simulations of QCD may 
be related to collective behaviour of QGP. Recently, very good fit 
to lattice results on EoS of QGP is also obtained \cite{ba.3} 
by treating QGP as 
SCQGP and again the nature of EoS is determined by plasma parameter which 
depends on collective properties of plasma.     
        
%\section{Acknowledgement:} 
\noindent 
{\bf Acknowledgement:} 

I thank the organizer and DAE of India for the financial help to attend 
the International conference on ICPAQGP at Kolkata, India, which 
inspired me to present this result.

\newpage
\begin{figure}
\caption { Plots of $P/ T^4 $ as a function of $T$ from 
our model and lattice results (symbols) for pure gauge (lower curve), 
2-flavor QGP (middle curve) and 3-flavor QGP (upper curve). } 
\label{fig 1}
\vspace{.75cm}

\caption { Plots of $\varepsilon/ T^4 $ as a function of $T/T_c$ from 
our model and lattice results (symbols) for pure gauge (lower curve), 
2-flavor QGP (middle curve) and 3-flavor QGP (upper curve). } 
\label{fig 2}
\vspace{.75cm}

\caption { Plots of $c_s^2$ as a function of $T/T_c$ from 
our model for pure gauge (lower curve), 
2-flavor QGP (middle curve) and 3-flavor QGP (upper curve) and also 
with lattice data for pure gauge. } 
\label{fig 3}
\vspace{.75cm}

\caption { Plots of $n_q/T^3$ as a function of $T/T_c$ from 
our model of 2-flavor QGP and also 
with lattice data (symbols)\cite{ak.1}. } 
\label{fig 4}
\vspace{.75cm}

\caption { Plots of $\Delta P/T^4$ as a function of $T/T_c$ from 
our model of 2-flavor QGP and also 
with lattice data (symbols). } 
\label{fig 5}
\vspace{.75cm}

\caption { Plots of $\chi_q /T^2$ as a function of $T/T_c$ from 
our model of 2-flavor QGP and also 
with lattice data (symbols). } 
\label{fig 6}
\vspace{.75cm}
\end{figure}
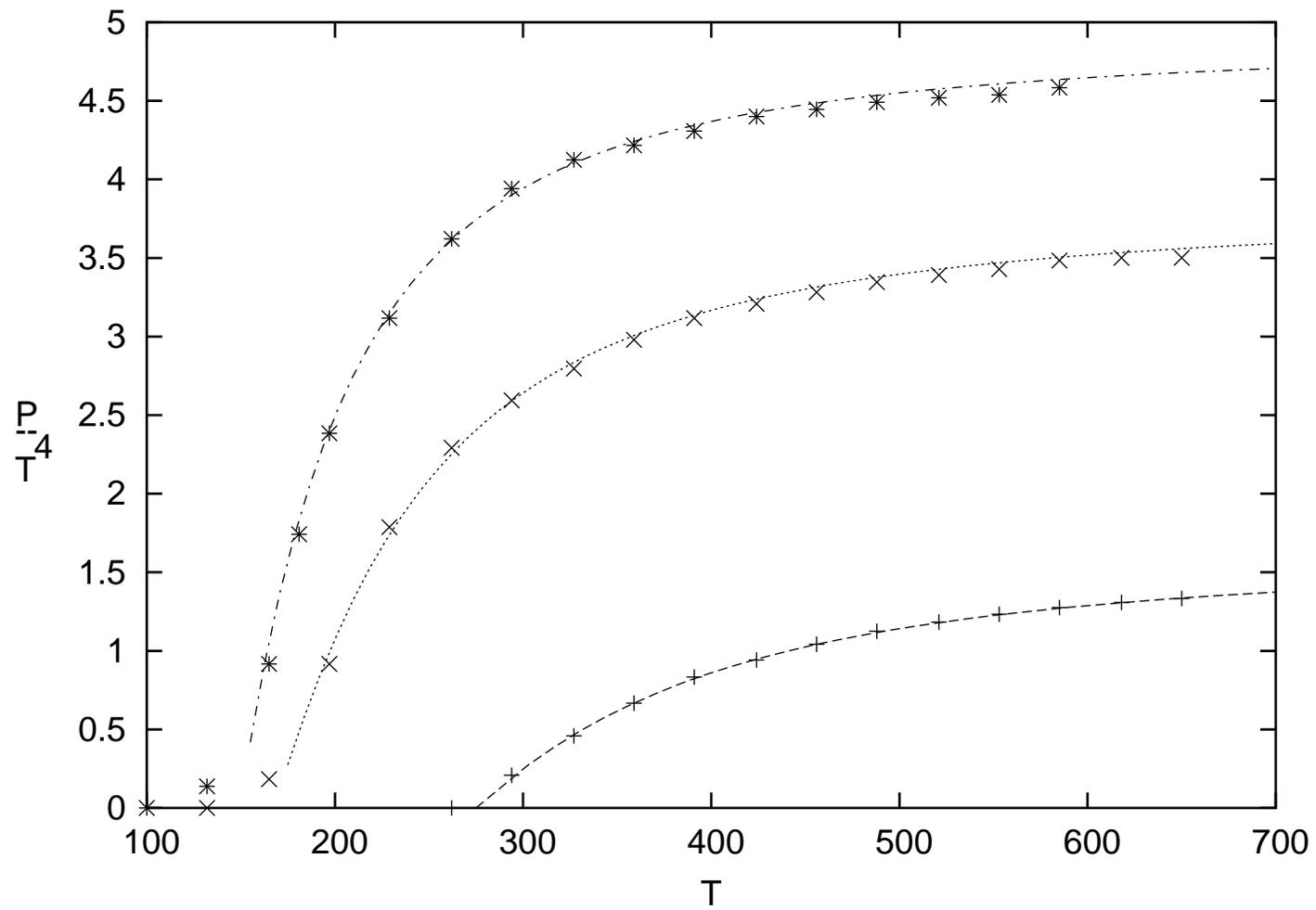
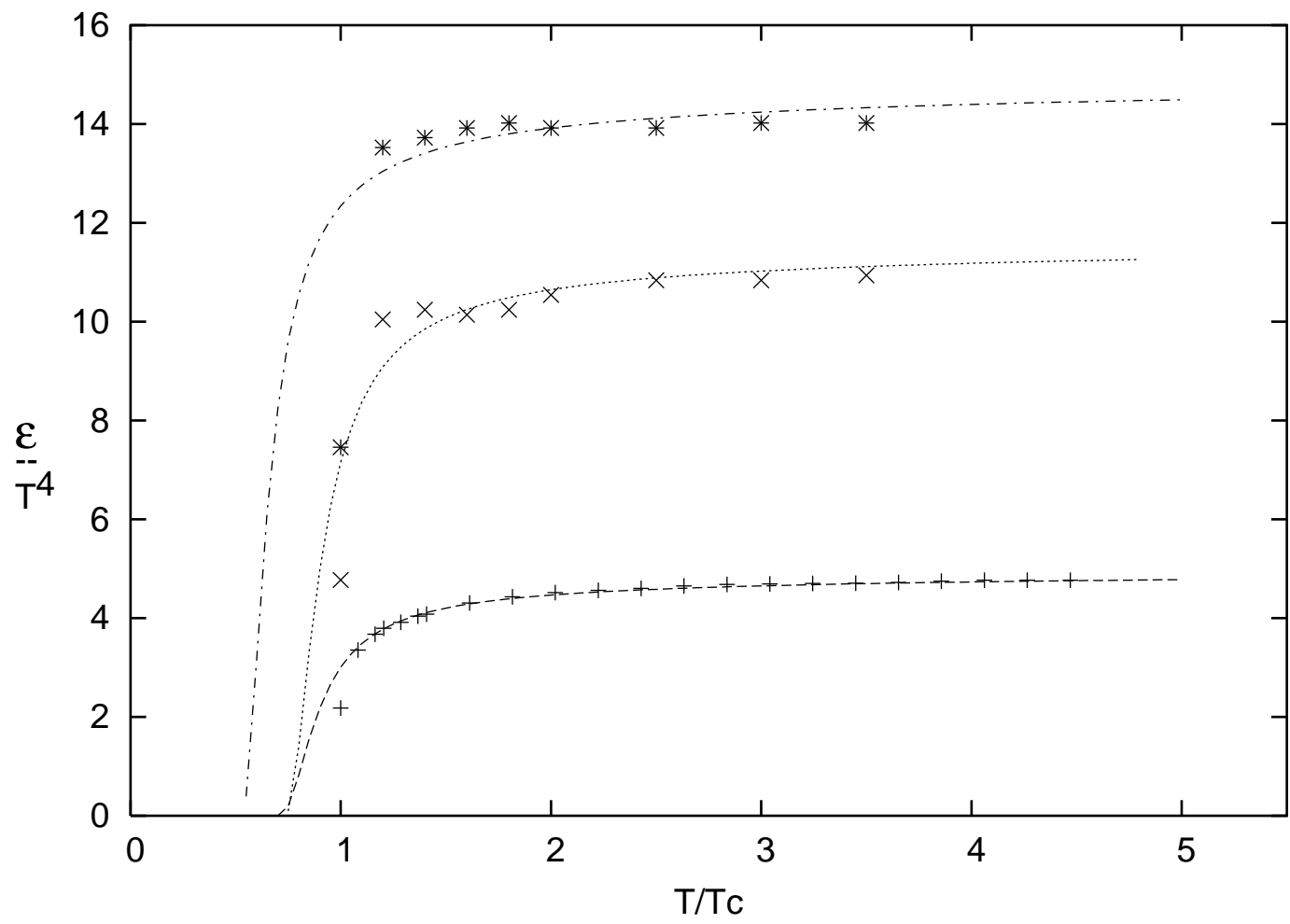
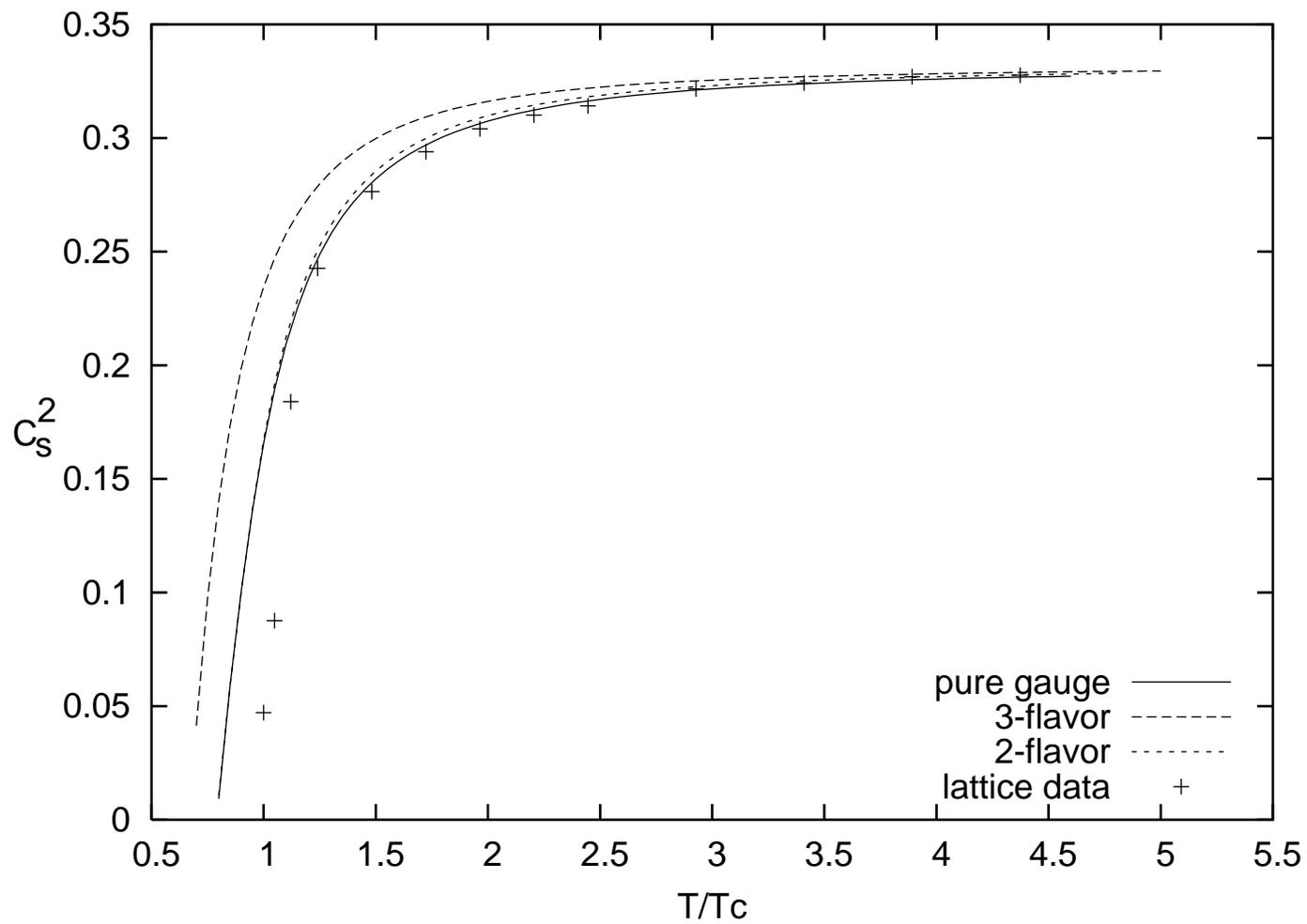
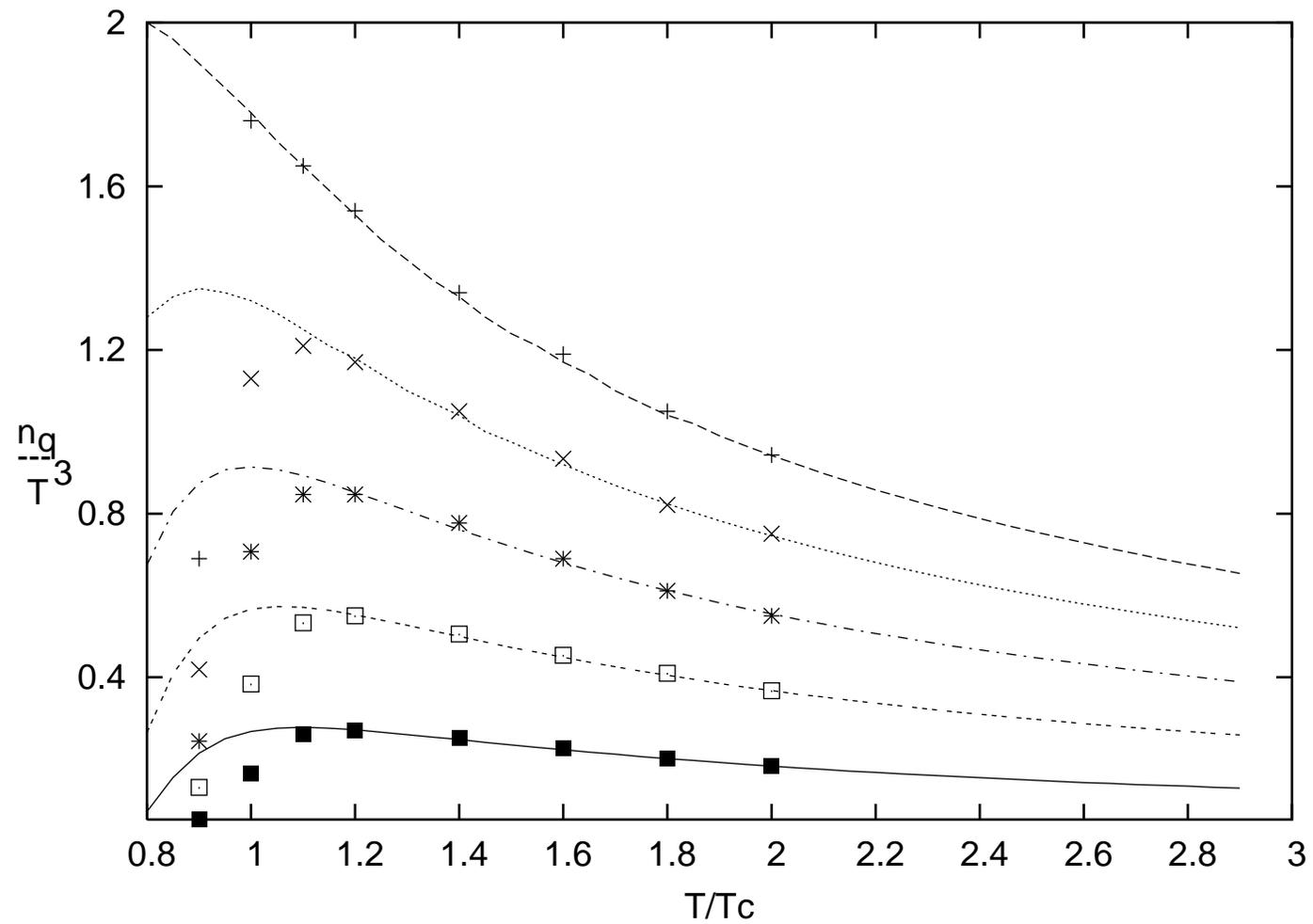
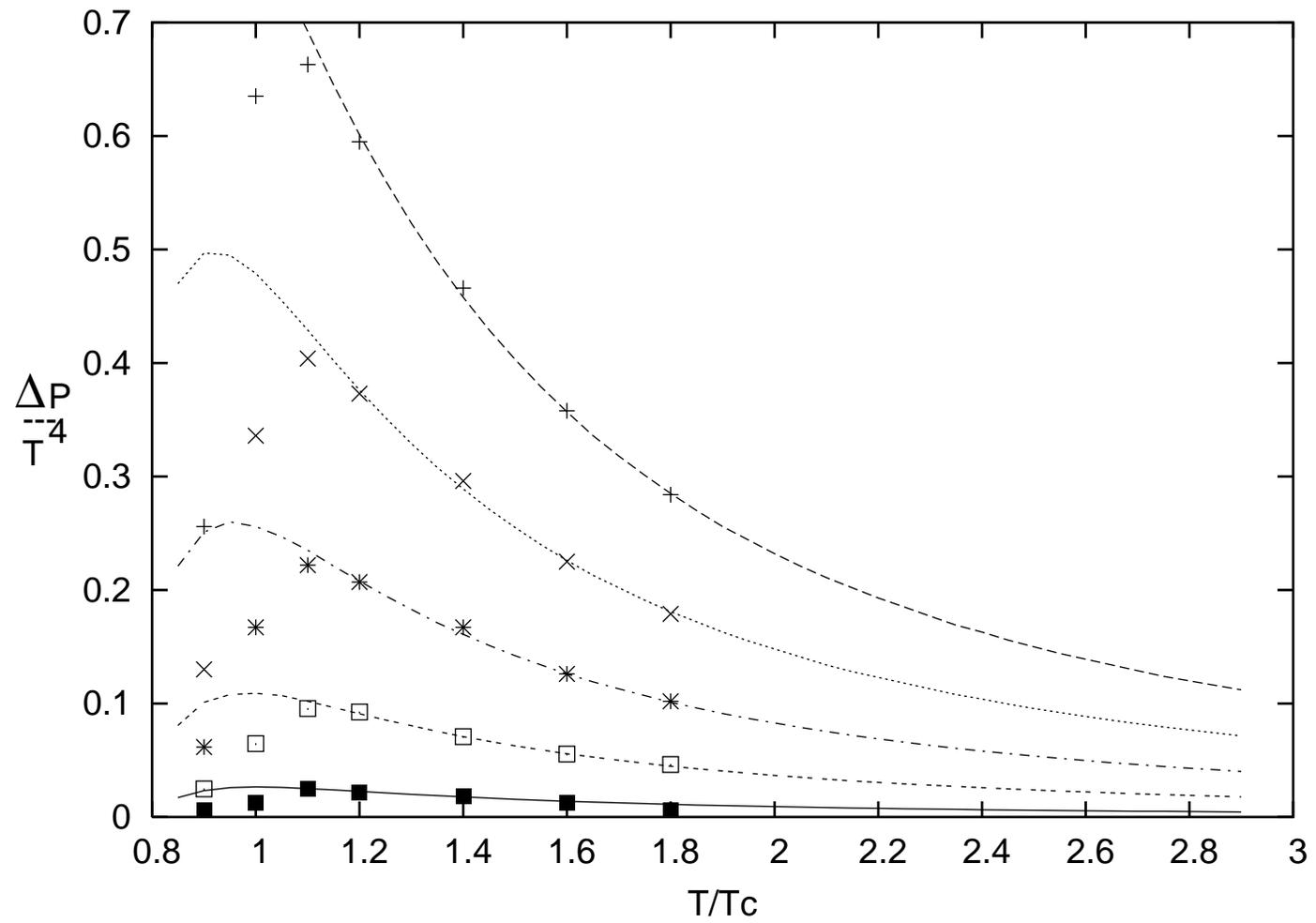
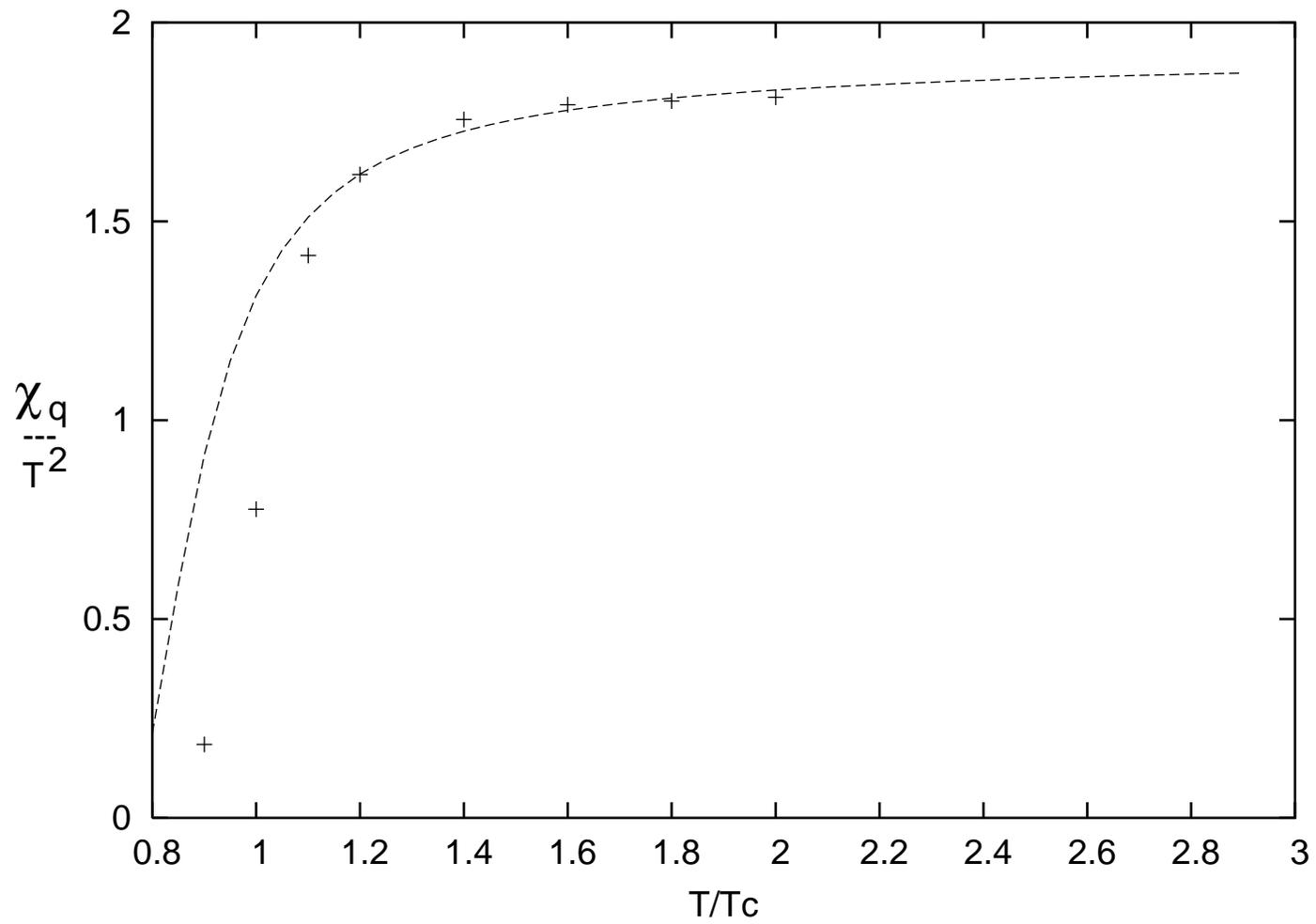
\end{document}